\documentclass[preprintnumbers,pre]{revtex4}
\usepackage{graphicx}% Include figure files
\usepackage{bm}% bold math
\begin{document}

\title{
Applications of Computer Simulations
and Statistical Mechanics in
Surface Electrochemistry}

\author{
P.A.\ Rikvold$^{1,2}$}\email{rikvold@scs.fsu.edu} 
\author{
I.~Abou Hamad$^{3}$}\email{ibrahim@hpc.msstate.edu}
\author{
T.~Juwono$^1$}\email{juwono@scs.fsu.edu}
\author{
D.T.\ Robb$^4$}\email{drobb@clarkson.edu}
\author{
M.A.\ Novotny$^{3,5}$}\email{man40@ra.msstate.edu}

\affiliation{
$^{1}$ 
Center for Materials Research and Technology,\\ 
School of Computational Science, and Department of Physics,\\
Florida State University, Tallahassee, FL 32306-4350, USA\\
$^{2}$ National High Magnetic Field Laboratory, Tallahassee, FL
32310-3706, USA\\
$^3$ HPC$^2$, Center for Computational Sciences, Mississippi State University,
Mississippi State, MS 39762-5167, USA\\
$^4$ Department of Physics, Clarkson University, Potsdam, NY 13699, USA\\
$^5$ Department of Physics \& Astronomy, Mississippi State University,\\
Mississippi State, MS 39762-5167, USA\\
}
%
% Use the package "url.sty" to avoid
% problems with special characters
% used in your e-mail or web address
%

\newcommand{\eqref}[1]{(\ref{#1})}
\newcommand{\mx}{\mathbf}
\newcommand{\wmx}{\mathbf{W}}
\newcommand{\state}{\mathbf{s}}
\newcommand{\initst}{\state (0)}
\newcommand{\finst}{\state {^*}}
\newcommand{\sgn}{\mathrm{sgn}}
\newcommand{\prob}{\mathrm{P}}
\newcommand{\stilde} {\tilde s}
\newcommand{\seta} {{\cal A}_t}
\newcommand{\setb} {{\cal B}_t}

\begin{abstract}
We present a brief survey of methods that utilize computer simulations
and quantum and statistical mechanics 
in the analysis of electrochemical systems.
The methods, Molecular Dynamics and Monte Carlo simulations and
quantum-mechanical density-functional theory, are
illustrated with examples from simulations of lithium-battery charging and
electrochemical adsorption of bromine on single-crystal silver electrodes. 
\end{abstract}
\maketitle

\section{Introduction}

The interface between a solid electrode and a liquid electrolyte is a 
complicated many-particle system, in which the electrode ions and electrons
interact with solute ions and solvent ions or
molecules through several channels
of interaction, including forces due to quantum-mechanical exchange, 
electrostatics, hydrodynamics,
and elastic deformation of the substrate. Over the last
few decades, surface electrochemistry 
\index{surface electrochemistry} has been revolutionized by new
techniques that enable atomic-scale observation and manipulation
of solid-liquid interfaces \index{solid-liquid interfaces} \cite{KOLB02,TANS06},
yielding novel methods for materials analysis, synthesis, and modification.
This development has been paralleled by equally revolutionary
developments in computer hardware and algorithms that by now enable
simulations with millions of individual particles \cite{VASH06}, so that
there is now significant overlap between system sizes that can be
treated computationally and experimentally. 

In this chapter, we discuss some of the methods available to study the
structure and dynamics of electrode-electrolyte interfaces using
computers and techniques based on quantum and 
statistical mechanics. These methods are 
illustrated by some recent applications.
The rest of the chapter is organized as follows. 
In Sec.~\ref{sec:LI}, we present fully three-dimensional, continuum
simulations by Molecular Dynamics \index{Molecular Dynamics} (MD)
of ion intercalation during charging of 
\index{Lithium-ion batteries} Lithium-ion batteries. 
In Sec.~\ref{sec:LG}, we discuss the simplifications that are possible by
mapping a chemisorption problem onto an effective lattice-gas (LG)
Hamiltonian \index{lattice gas}, 
and in Sec.~\ref{sec:DFT} we demonstrate how input parameters
for a statistical-mechanical LG model can be estimated from quantum-mechanical
\index{density-functional theory} density-functional theory (DFT) calculations.
Section~\ref{sec:MC} is devoted to a discussion of 
\index{Monte Carlo simulations} Monte Carlo (MC)
simulations, both for equilibrium problems (Sec.~\ref{sec:EMC}) and for
dynamics (Sec.~\ref{sec:KMC}). As an example of the latter, we present in
Sec.~\ref{sec:FORC} a simulational demonstration of a method to classify
surface phase transitions in adsorbate systems, 
which is an extension of standard cyclic
voltammetry (CV): the \index{First-order Reversal Curve} 
Electrochemical First-order Reversal Curve
(EC-FORC) method. A concluding summary is given in Sec.~\ref{sec:CONC}.

\section{Molecular Dynamics Simulations of Ion
Intercalation in Lithium Batteries}
\label{sec:LI}

The charging process in Lithium-ion batteries is marked by the intercalation
of Lithium ions into the graphite anode material. Here we present MD
simulations of this process and suggest a new charging method that has 
the potential for shorter charging times, as well as the possibility of 
providing higher power densities.
\begin{figure}[t]
\vspace{0.1truecm}
\begin{center}
\includegraphics[angle=0,height=.45\textwidth]{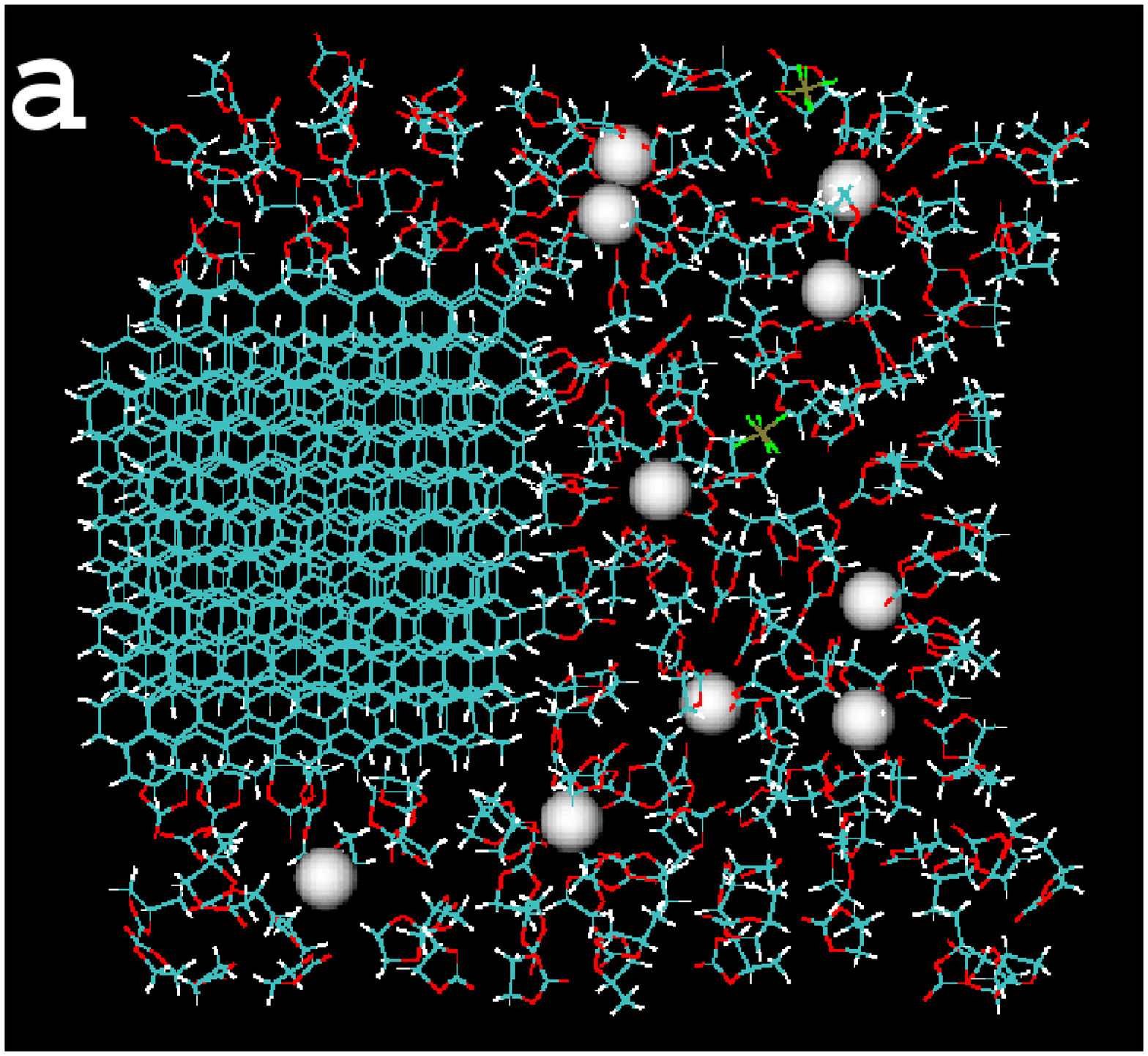}
\hspace{0.01truecm}
\includegraphics[angle=0,height=.45\textwidth]{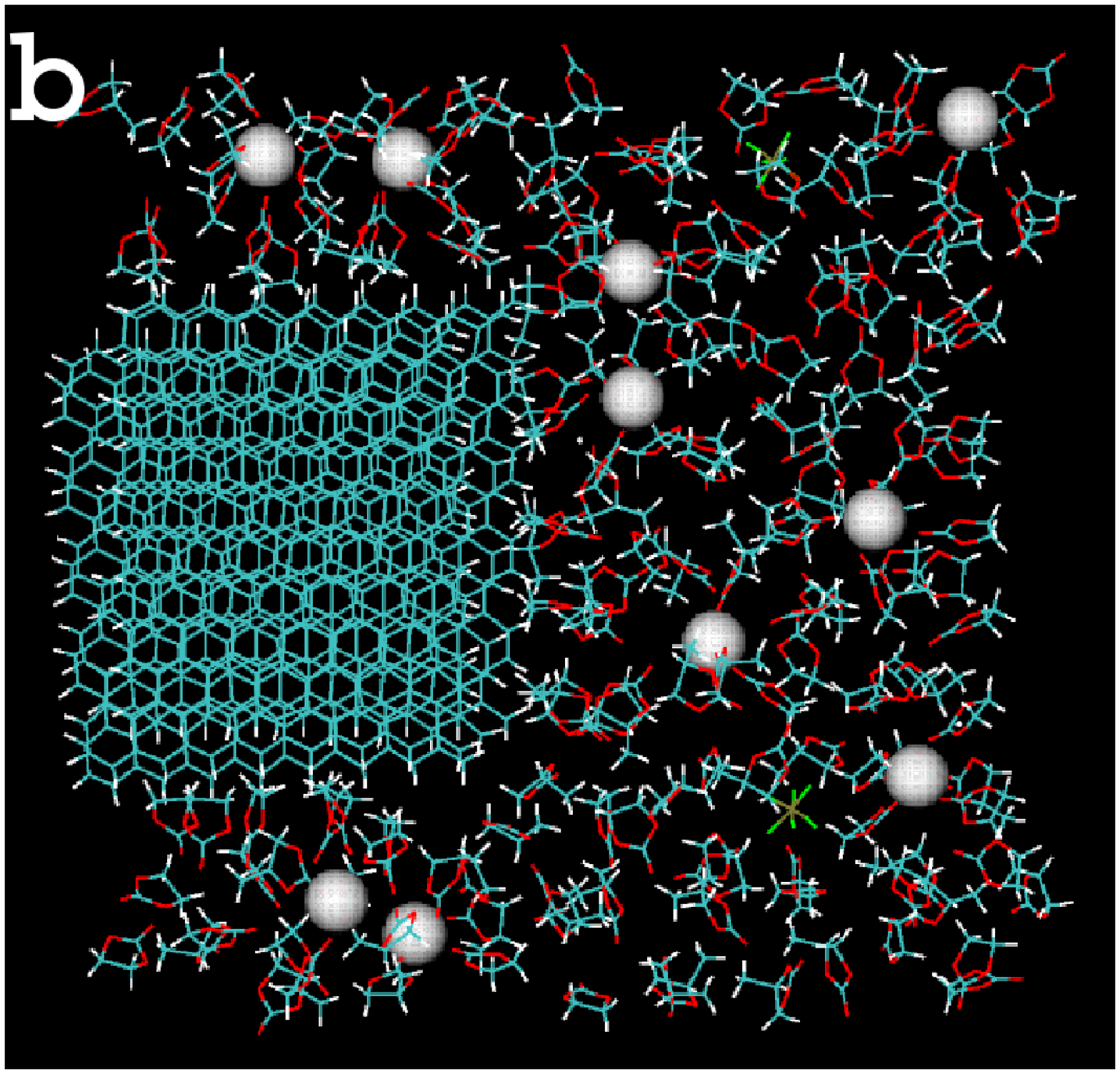}\\
\vspace{0.1truecm}
\end{center}
\caption[]{
({\bf a\/})~Snapshot of the model system containing four graphite
sheets,
two PF$_{6}^{-}$ ions and ten Li$^{+}$ ions (spheres), 
solvated in $69$ propylene
carbonate and $87$ ethylene carbonate molecules after reaching constant
volume in the $NPT$ ensemble.
({\bf b\/})~Snapshot after $200$~ns MD simulation. The ensemble is the $NVT$
ensemble. The system has periodic boundary conditions and is simulated
at one atm and 300~K. Top view, perpendicular to the plane of the
graphite sheets.
}
\label{fig:MDsystem}
%\vspace*{-0.40cm}
\end{figure}

\subsection{Molecular Dynamics and Model System}
\label{sec:MDMS}

Molecular Dynamics is based on solving the classical equations of motion
for a
system of $N$ atoms interacting through forces derived from a
potential-energy function
\cite{Allen:92,Haile:92,Rapaport:04,VOTE97,VOTE97B}. 
From the potential energy $E_{\rm P}$,
the force on the $i$th atom, $F_{i}$, is calculated. Thus, the equation of
motion is 
\begin{equation}
\label{eq:eqmot}
F_{i}(t)=-\frac{\partial E_{\rm P}}{\partial r_{i}}=m_{i}
\frac{\partial v_{i}}{\partial t}=m_{i}\frac{\partial^2 r_{i}}{\partial t^2}
\end{equation}
where $r_{i}$, $v_{i}$, and $m_{i}$ are the position,
velocity, and mass of the $i$th atom, respectively.
Consequently, the quality of the
simulations strongly depends on the ability of the classical force field
to reasonably describe the atomistic behavior.

The newly developed General Amber Force Field (GAFF)~\cite{GAFF:04} was
used to approximate the bonded interactions of all the simulation molecules,
while the simulation package Spartan (Wave-function, Inc., Irvine, CA) was
used at the Hartree-Fock/6-31g* level to obtain the necessary point charges for
each of the atoms. To simulate a charging field, the charge on the carbon
atoms of
the graphite sheets was set to $-0.0125$~\textit{e} per atom. The bonded
(first three terms of Eq.~(\ref{eq:AMBER})) and non-bonded (last term)
interactions
in the AMBER Force Field are represented by the following potential-energy 
function:
%\begin{eqnarray}
\begin{equation}
\label{eq:AMBER}
E_{\rm P} = \sum_{\rm bonds} K_r (r - r_{eq})^2
                     + \sum_{\rm angles} K_\theta (\theta -
		     \theta_{eq})^2
              + \sum_{\rm dihedrals} {V_n \over 2}
                                       [1 + {\rm cos}(n\phi - \gamma)] 
             + \sum_{i<j} \left [ {A_{ij} \over
	     R_{ij}^{12}} -
                                          {B_{ij} \over R_{ij}^6} +
                                          {q_iq_j \over \epsilon R_{ij}}
                                 \right ]
%\end{eqnarray}
\end{equation}
where $K_{\rm r}$, $K_{\theta}$ and $V_n$ are the bond stretching, bending
and torsional constants respectively, the constants $A$ and $B$ define
van der Waals' interactions between unbonded atoms, 
and $\epsilon$ is the electrostatic permittivity. The simulation
package NAMD~\cite{NAMD} was used for the MD simulations, while
the graphics package 
VMD~\cite{VMD} was used for visualization and analysis of the simulation
results.

The model system representing the anode half-cell is composed of four
graphite
sheets (anode) containing $160$ carbon atoms each, two PF$_{6}^{-}$
ions, and
ten Li$^{+}$ ions, solvated in an electrolyte made of $69$ propylene
carbonate and $87$ ethylene carbonate molecules 
(see Fig.~\ref{fig:MDsystem}({a})). The
graphite sheets were fixed from one side by keeping the positions of the
edge carbon atoms fixed.
\begin{figure}[t]
\vspace{0.8truecm}
\begin{center}
\includegraphics[angle=0,width=0.85\textwidth]{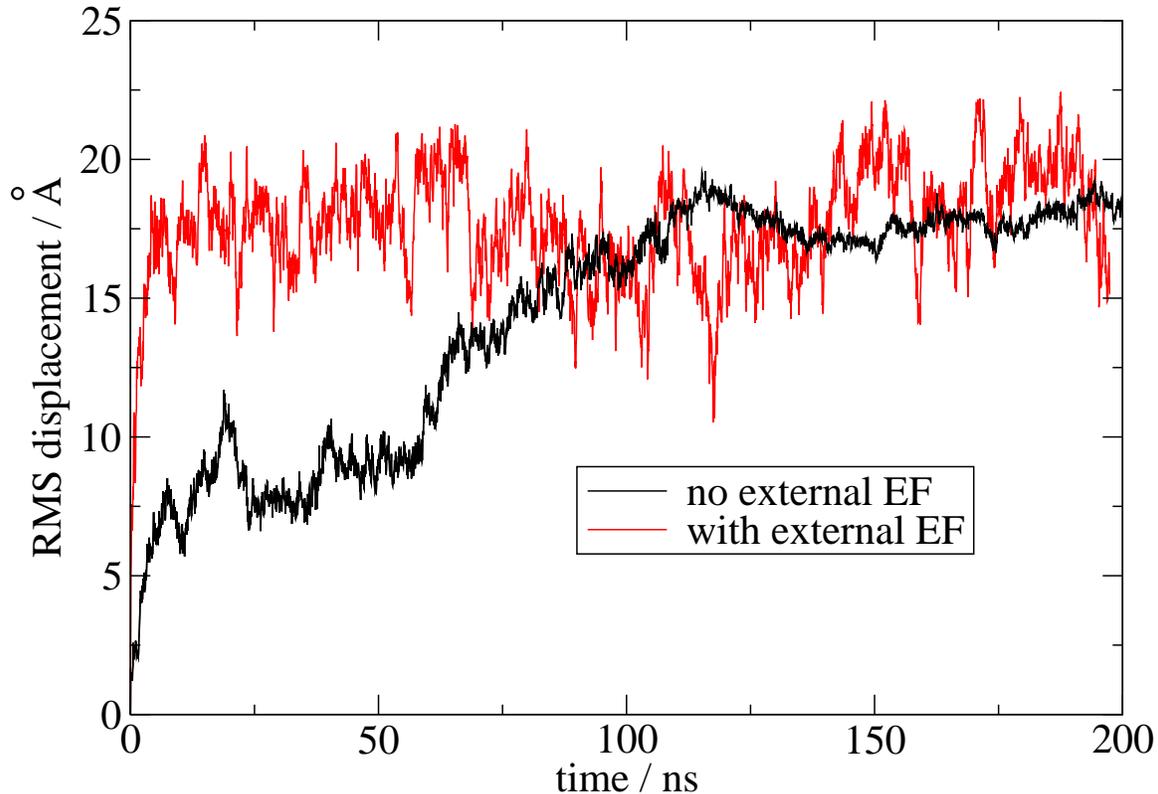}
\end{center}
\caption[]{\small
Root-mean-square displacement of Lithium ions as a function of time.
Diffusion is much faster with the additional oscillating electric field
(amplitude 5 kCal/mol, frequency 25 GHz).
}
\label{fig:rmsd}
\end{figure}

\subsection{Simulations and Results}
\label{sec:MDSR}
After energy minimization, the simulations were run at constant pressure
using
a Langevin piston Nos\'{e}-Hoover method~\cite{Martyna:94,Feller:95} as
implemented in the NAMD software package until the system has reached
its
equilibrium volume at a pressure of $1$ atm and 300~K in the $NPT$
(constant particle number, pressure, and temperature) 
ensemble. The system's behavior was then simulated for $200$~ns ($100$
million steps)
in the $NVT$ (constant particle number, volume, and temperature)
ensemble. Two observations were made: first, the Li$^{+}$ ions stayed
randomly distributed within the electrolyte, and second, none of the
Li$^{+}$ ions had intercalated between the graphite sheets after $200$~ns
(see Fig.~\ref{fig:MDsystem}({b})).

While the Lithium ions do not intercalate within the simulation time
given
above, it is expected that given enough time they will move towards the
graphite sheets and get intercalated. To test whether intercalation is
possible
in such a model system, one of the Lithium ions was positioned between the
graphite sheets at the beginning of a simulation, and we observed whether
it diffused out from between the sheets. The Lithium ion stayed
intercalated, even after $400$~ns.

In order for intercalation to occur, the Lithium ion has first to
diffuse within
the electrolyte until it reaches the graphite electrode. Consequently,
faster
diffusion would result in faster intercalation and shorter charging
time.
In order to increase the diffusion of Lithium ions in the electrolyte, 
we explored a new charging method. 
In addition to the charging field due to the fixed charge
on the graphite carbons, an external oscillating square-wave field
(amplitude
5~kCal/mol, frequency 25~GHz) was applied in the direction perpendicular
to the plane of the graphite sheets. Not only does this additional field
increase
diffusion, but also some of the Lithium ions intercalate into the
graphite
sheets within an average time of about 50~ns. Figure~\ref{fig:rmsd} shows a
plot of the root-mean-square displacement of Lithium ions as a
function of time for a system with and without an applied external field.
The increased diffusion and intercalation indicate that a charging
protocol involving an oscillating field may decrease the charging
time and possibly increase the battery's power density.

\section{Lattice-gas Models of Chemisorbed Systems}
\label{sec:LG}

As mentioned in the Introduction,  
even the simplest electrosorption systems are extremely complicated. 
This complexity means that a comprehensive theoretical
description that enables predictions for phenomena on macroscopic scales
of time and space is still generally impossible
with present-day methods and technology.
(Note that MD simulations, such as those presented in Sec.~\ref{sec:LI},
are only possible up to times of a few hundred nanoseconds.)
Therefore, it is necessary to use a variety of analytical and
computational methods and to study various 
simplified models of the solid-liquid interface. 
One such class of simplified models are {\it Lattice-gas\/} (LG) models, in
which chemisorbed particles (solutes or solvents) can only be located at
specific adsorption sites, commensurate with the substrate's crystal
structure. This can often be a very good approximation, as for instance
for halides on the (100) surface of Ag, for which it can be shown that
the adsorbates spend the vast majority of their time near the four-fold
hollow surface sites \cite{MITC02}. 
A lattice-gas approximation to such a continuum model,
appropriate for chemisorption of small molecules or ions
\cite{HUCK90,BLUM94A,BLUM96,GAMB93B,RIKV95,JZHA95B},
is defined by the discrete, effective grand-canonical Hamiltonian,
\begin{equation}
{\cal H}_{\rm LG}
= \sum_n [ -\Phi^{(n)}
\sum_{\langle ij \rangle}^{(n)} c_i c_j ]
+ {\cal H}_3 - \bar{\mu} \sum_i c_i
\; .
\label{eq:LG}
\end{equation}
Here, the lattice sites $i$ are the preferred adsorption sites (the
minima of the continuous corrugation potential),
and $c_i$ is a local occupation variable, with 1
corresponding to an adsorbed particle and 0 to a solvated site.
The sums $\sum_{\langle ij \rangle}^{(n)}$ and $\sum_i$ run over all
$n$th-neighbor pairs and over all adsorption sites,
respectively, $\Phi^{(n)}$ is the effective $n$th-neighbor
pair interaction, and $\sum_n$ runs over the interaction ranges.
The term ${\cal H}_3$ contains multi-particle interactions
\cite{EINS91,HYLD05,STAS06}.
The sign convention is such that $\phi < 0$ implies
repulsion, and $\bar{\mu}>0$ favors adsorption. 
Equation~(\ref{eq:LG}) is also easily generalized to multiple
species \cite{RIKV88B,COLL89}.

To connect the electrochemical potentials to the concentrations
in bulk solution of species X, [X],
and the electrode potential, $E$, one has
(in the dilute-solution approximation)
\begin{equation}
\label{eq2}
\bar{\mu}_{\rm x}(T,[{\rm X}],E)
= \bar{\mu}_{\rm x}^0
+ k_{\rm B} T \ln ( [{\rm X}] / [{\rm X}]^0 )
- e \int_{E^0}^E \gamma_{\rm x}(E') {\rm d}E' \;,
\end{equation}
where $k_{\rm B}$ is Boltzmann's constant,
$T$ the temperature, $e$ the elementary charge,
and $\gamma_{\rm x}(E)$ the electrosorption valency
\cite{SCHM96,VETT72A,VETT72B,RIKV07} of X.
The importance of the integral over the potential-dependent
electrosorption
valency (rather than just the product $e\gamma_{\rm x}(E)E$)
analogous to the case of potential-independent $\gamma_{\rm x}$)
was pointed out in Ref.~\cite{HAMA05B}.
The quantities superscripted ``0'' are reference values
that include local binding energies.
The interaction constants and electrosorption valencies are {\it
effective\/} parameters influenced by several physical
effects, including electronic structure \cite{EINS91,HYLD05,STAS06},
surface deformation,
(screened) electrostatic interactions \cite{KOPE98,GLOS93A,GLOS93B},
and the fluid electrolyte \cite{BLUM90,IGNA98}.
The density conjugate to $\bar{\mu}_{\rm x}$ is
the  coverage relative to the number $N$ of adsorption sites, 
\begin{equation}
\Theta_{\rm X} = N^{-1} \sum_i c_i
\;.
\label{eq:Theta}
\end{equation}

\section{Calculation of Lattice-gas Parameters by Density Functional Theory}
\label{sec:DFT}

There are many methods to estimate lattice-gas parameters. One of these
is comparison of MC simulations (see Sec.~\ref{sec:MC}) of a LG model with
experimental adsorption isotherms. For detailed descriptions of this
method we refer to 
Refs.~\cite{KOPE98,BROW99A,MITC00A,MITC00C,HAMA03,HAMA05B}. 
Here we instead concentrate on the purely theoretical method based
on quantum-mechanical DFT calculations \cite{STAS06}. 

DFT is the most widely used
method to calculate ground-state properties of many-electron systems.
It is based on the Hohenberg-Kohn theorem, which states that all
properties of the many-particle ground state can be expressed in terms
of the ground-state electron charge-density distribution
\cite{Hohenberg} and leads to the Kohn-Sham equations for
single-particle wave functions \cite{Kohn}.
%\begin{equation}
%\left(
%%-\frac{1}{2}\bigtriangledown^2+V_{\rm ext}+V_{\rm class}+V_{\rm xc}
%\right)\phi_n
%= \varepsilon_n\phi_n \;.
%\label{eq-2}
%\end{equation}
These are second-order differential equations, which include potential
terms due to the ions and the classical Coulomb repulsive energy between
the electrons, as well as the electronic exchange-correlation energy,
and they are solved self-consistently. For surface structural studies, DFT
is usually performed using pseudopotentials with
slab models and plane-wave basis sets.
The slab consists of a finite number of atomic layers,
periodic in the direction parallel to the surface, which can either be
repeated periodically in the third direction
(separated by a vacuum interval), or not. 
The fluid solvent can be considered either as an effective continuum,
%such as in the Self-Consistent Reaction Field (SCRF) methods
%\cite{Cramer,Tomasi,Mennucci,Cossi1,%
%Cances1,Barone1,Adamo,Barone2,Cances2,Wiberg,Foresman,Cossi2,%
%Wong1,Wong2,Wong3,Wong4,Miertus1,Miertus2},
or by molecular models.  

\begin{figure}[t]
\vspace{-0.4truecm}
\begin{center}
\includegraphics[angle=0,width=.75\textwidth]{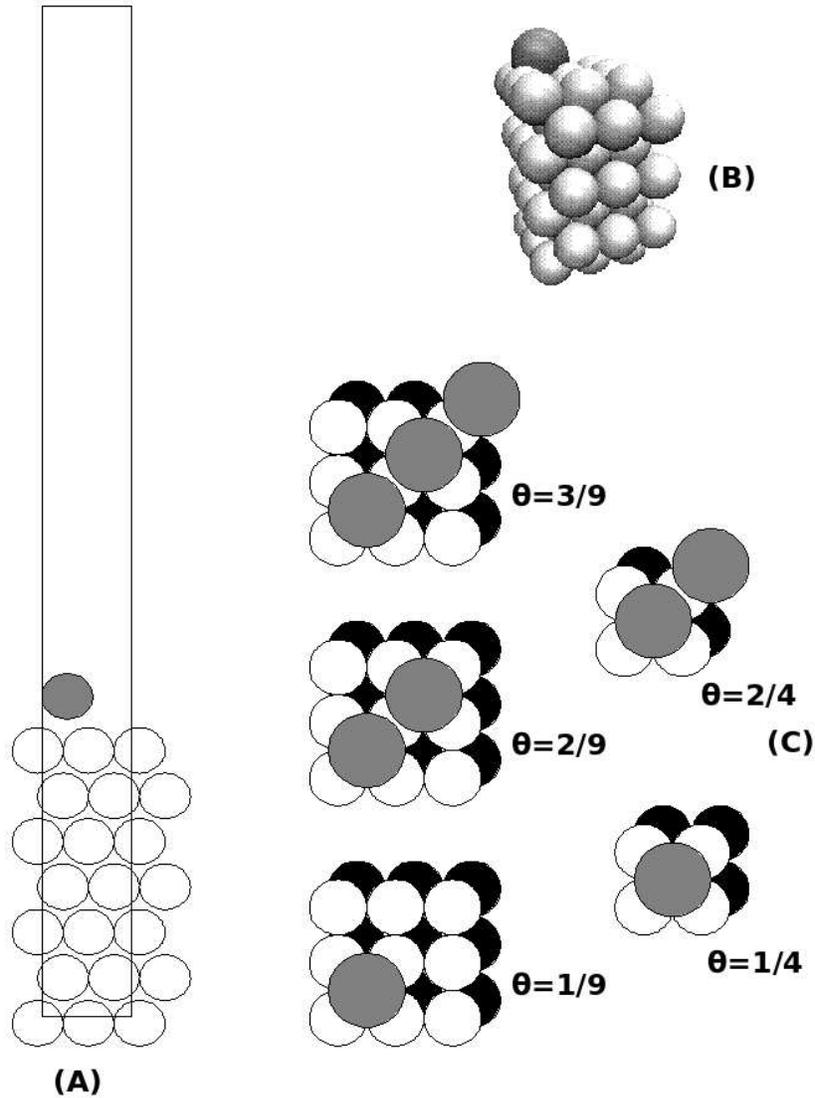}
\end{center}
\caption[]{
({\bf A\/}) 
Cross section of a $3 \times 3$ supercell with $ \Theta = 1/9$. 
({\bf B})
Three-dimensional representation of the same cell and coverage.
({\bf C}) 
Top view of a $ 3 \times 3 $ surface and a $ 2 \times 2 $ surface with various
coverages.
}
\label{fig:cell}
%\vspace*{-0.40cm}
\end{figure}
Here we present preliminary results on a DFT calculation of lateral
interaction constants pertaining to 
a lattice-gas model for the adsorption of Br
on single-crystal Ag(100) surfaces \cite{HAMA03,HAMA04,RIKV07,MITC00A,MITC00C}.
The lattice-gas model is represented by Eq.~(\ref{eq:LG}) on a square
lattice with lattice constant $a = 2.95$~\AA, $\mathcal{H}_3 = 0$, 
infinitely repulsive interactions for adparticles at nearest-neighbor
sites, and the long-range repulsion 
\begin{equation}
\phi_{ij}
%=\frac{R_{\rm nnn}^3}{R_{ij}^3}\phi_{\rm nnn}
=\frac{(\sqrt{2})^3}{r_{ij}^3}\phi_{\rm nnn} 
\;\;\; {\rm for} \;\;\; r_{ij} \ge \sqrt{2}
\;,
\label{eq:phi}
\end{equation}
which is compatible with dipole-dipole interactions or elastically
mediated interactions. (Here, $r_{ij}$ is given in units of $a$.)
Since the DFT calculations are performed in the
canonical ensemble (fixed adsorbate coverage), $\bar{\mu}$ in 
Eq.~(\ref{eq:LG}) is replaced by the binding energy 
of a single adparticle, $E_{\rm b}$. 

We prepared slabs with seven metal layers, which were
placed inside a supercell with periodic boundary conditions. Two
different sizes of supercells were used:
a $2 \times 2$ supercell with the size of $2a \times 2a \times 36.95$~\AA, 
and a  $3 \times 3$ supercell with the size of $3a \times 3a \times 36.95$~\AA. 
The vacuum region above the surface was twice the thickness of the slab,
and the orientation of the surface normal was in the $z$ direction. One,
two, and three Br atoms were placed on the $3 \times 3$ surface to
represent coverages $\Theta= 1/9$, $2/9$, and $1/3$. Two Br 
atoms were placed on the $2 \times 2$ surface to represent $\Theta=1/2$, 
and one to represent $\Theta = 1/4 $. 
Supercells with different coverages of Br are shown in
Fig.~\ref{fig:cell}.

The DFT calculations were performed using the Vienna Ab Initio
Simulation Package (VASP) \cite{KRES93,KRES96A,KRES96B}. The basis
set was plane-wave, with the generalized gradient-corrected
exchange-correlation function \cite{Perdew2,Perdew1}, and Vanderbilt
pseudopotentials \cite{Vanderbilt}.
%, and a cut-off energy of 400 eV. 
The $k$-point mesh was generated using the
Monkhorst method \cite{MONK76} with a $5 \times 5 \times 1$ grid for the
$3 \times 3$ cells and a $7 \times 7 \times 1$ grid for the $2 \times 2$
cells. All calculations were done on a $54 \times 54 \times 192$
real-space grid. 

Individual DFT calculations provide total energies, $E$,
and charge densities,  $\rho(\vec{x})$. 
The adsorption energy $E_{\rm ads}$ for a single adatom
and the corresponding charge-transfer 
function $\Delta \rho(\vec{x})$ are obtained from calculations of
the adsorbed system and isolated slab and atoms as follows:
\begin{equation}
E_{\rm ads}=\left[E_{\rm syst}-E_{\rm slab}
\right] / N_{\rm ads} -E_{\rm Br} 
\label{eq:Eads}
\end{equation}
and \cite{MITC04}
\begin{equation}
\Delta\rho(\vec{x}) = \lbrack \rho(\vec{x})_{\rm syst} 
-\rho(\vec{x})_{\rm slab}
\rbrack/N_{\rm ads} - \rho(\vec{x})_{\rm Br} \;,
\end{equation}
where $N_{\rm ads} = N \Theta$ is the number of adsorbed Br atoms in the
cell, and the quantities subscripted Br refer to a single, isolated Br
atom. 

\begin{figure}[t]
\vspace{0.8truecm}
\begin{center}
\includegraphics[angle=0,width=.85\textwidth]{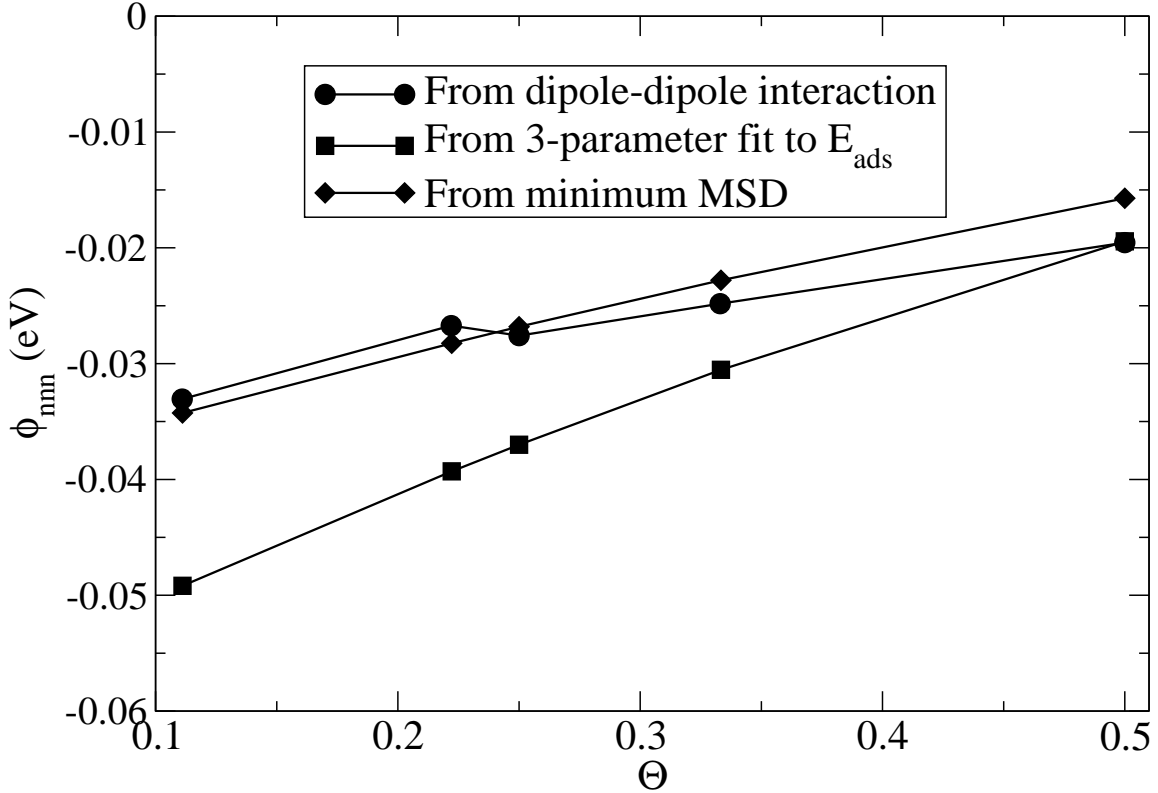}
\end{center}
\caption[]{
Three different estimates of the LG interaction constant $\phi_{\rm nnn}$. 
Circles: 
Based on Eq.~(\protect\ref{dip_dip}) with the dipole moment $p$ directly
obtained from the DFT calculation. 
Squares:
Based on a three-parameter fit to the DFT adsorption energy $E_{\rm ads}$
as described in Eq.~(\protect\ref{eq:fit1}). 
Diamonds:
Based on minimizing mean-square deviations (MSD)
from the estimate based on the DFT dipole
moment $p$, constrained to retain
a low value of the $\chi^2$ from the fit to $E_{\rm ads}$. 
See discussion in the text. }
\label{fig:DFTres}
%\vspace*{-0.40cm}
\end{figure}

Since the system is electrically neutral, the integral over space of 
$\Delta\rho(\vec{x})$ vanishes. The surface dipole moment is defined as
\begin{equation}
p= \int z \Delta \rho(z)dz \;.
\label{surface_dip}
\end{equation}
Kohn and Lau \cite{KOHN76} have shown that the non-oscillatory part of the
dipole-dipole interaction energy between adsorbates separated by a
distance $R$  behaves as
\begin{equation}
\phi_{\rm dip-dip}=\frac{2p_ap_b}{4\pi\epsilon_0R^3}
\label{dip_dip}
\end{equation}
for large $R$ (in our case larger than the nearest-neighbor distance).
This result is twice what one might na{\"\i}vely expect. 
Thus, the next-nearest-neighbor interaction constant from 
Eq.~(\ref{eq:phi}) would be  
\begin{equation}
\phi_{\rm{dip-dip~nnn}}=\frac{2p^2}{4\pi\epsilon_0R^3_{\rm{nnn}}}
\label{dftdip}
\end{equation}
with $p$ obtained from the DFT by Eq.~(\ref{surface_dip}).
This estimate, which depends on $\Theta$,
is included in Fig.~\ref{fig:DFTres} as solid circles. 

Alternatively, the interaction constant $\phi_{\rm nnn}$ in the LG
Hamiltonian, Eq.~(\ref{eq:LG}), can be estimated by performing a 
nonlinear least-squares fit of the $\Theta$-dependent DFT adsorption energy 
$E_{\rm ads}$ in Eq.~(\ref{eq:Eads}) to 
\begin{equation}
E_{\rm ads} = - \phi_{\rm nnn} \Sigma_\Theta - E_{\rm b} \Theta
\label{eq:fit1}
\end{equation}
with $\phi_{\rm nnn} = A (1 + B \Theta)^2$, using the three fitting
parameters $A$, $B$, and $E_{\rm b}$. This is consistent with the
theoretical prediction of Eq.~(\ref{dftdip})
with a dipole moment that depends linearly on $\Theta$. The quantity 
\begin{equation}
\Sigma_{\Theta}=\frac{(\sqrt{2})^3}{N} 
\sum_{i<j} \frac{c_ic_j}{r_{ij}^3} 
\label{sigmatheta}
\end{equation}
can be calculated numerically to any given accuracy for a particular
coverage and adsorbate configuration. 
This estimate for $\phi_{\rm nnn}$ 
is included in Fig.~\ref{fig:DFTres} as solid squares. 
It does not agree particularly closely with the result obtained from the
dipole moments. However, we found that the $\chi^2$ of the fit,
considered as a function of the
fitting parameters, was characterized by an
extremely wide and shallow basin surrounding its minimum.  
We therefore further minimized the mean-square deviation (MSD)
between the values of $\phi_{\rm nnn}$ obtained from this fitting procedure and
those obtained directly from Eq.~(\ref{dftdip}) with the DFT values for
$p$ {\it within the
three-dimensional parameter region for which the original $\chi^2$ was
close to its minimum\/}. This procedure gave significantly improved  
consistency between the two estimates for $\phi_{\rm nnn}$, 
without a significant increase in $\chi^2$. The final result is
shown as solid diamonds in Fig.~\ref{fig:DFTres}, and the corresponding
parameters are listed in Table~\ref{tab:fits}. 

The average value of $\phi_{\rm nnn}$ obtained by this method is
consistent with that found by fitting equilibrium MC simulations (see
Sec.~\ref{sec:EMC}) to experimental adsorption isotherms in aqueous
solution (approximately $-21$~meV). However, no
significant coverage dependence was found in the analysis of the
experimental data \cite{HAMA03,HAMA05B}.
It is not surprising that results from {\it in situ\/}
experiments and {\it in vacuo\/} DFT calculations should show some
differences, and we find it encouraging that the average results are
consistent. Application of the method described here to Cl/Ag(100) gave less
consistent results than for Br, possibly indicating that the effective
interactions for Cl are not purely dipole-dipole in nature \cite{JUWO08}. 
\begin{table}[t]
\begin{center}
\caption{
Results for the fits of the $\Theta$-dependent lattice-gas interaction 
constant $\phi_{\rm nnn}$ according to the two methods described in the
text. Here, $\nu$ is the number of degrees of freedom (number of data
points minus number of parameters, here equal to 2) for the initial nonlinear
least-squares fit of $\phi_{\rm nnn}$ to the DFT adsorption energy 
$E_{\rm ads}$, while MSD is the mean-square deviation between this
estimate and the estimate obtained directly from the DFT dipole moment. 
Minimizing MSD within the basin of low $\chi^2$ significantly reduces
MSD (see the greatly improved agreement in 
Fig.~\ref{fig:DFTres}) without significantly increasing $\chi^2$. 
}
\label{tab:fits}
\begin{tabular}{cccccc}
\\
\hline
\\
Method & $A$ & $B$ & $E_{\rm b}$ & $\chi^2/\nu$ & MSD/$\nu$ \\
\hline
\hline
\\
Min.\ $\chi^2$~ & ~$-6.017\times 10^{-2}$~ & ~$-0.8632$~ 
& ~3.102~ & ~$2.362\times 10^{-5}$~ &
~$1.803\times 10^{-4}$~\\
Min.\ MSD~      & ~$-4.085\times 10^{-2}$~ & ~$-0.7595$~ 
& ~3.070~ & ~$2.675\times 10^{-5}$~ &
~$7.692\times 10^{-6}$~\\
\\
\hline
\end{tabular}
\end{center}
\end{table}

\section{Monte Carlo Simulations}
\label{sec:MC}

\subsection{Equilibrium Monte Carlo}
\label{sec:EMC}

As a method to obtain equilibrium properties of a system described by a
particular Hamiltonian, MC is more accurate than
mean-field approximations, especially for low-dimensional
systems near phase transitions \cite{BROW99A,LAND00}. This is an effect
of fluctuations which, while ignored or underestimated by mean-field
methods, are very important in two-dimensional systems.
Given the rapid evolution of computers
and the relative ease of programming of MC codes, this
is our method of choice for equilibrium and dynamic
studies of both lattice-gas and continuum models.

The goal of an equilibrium MC code is to bring the system to
equilibrium as rapidly as possible, and then sample the equilibrium
distribution as efficiently as possible. The only requirement is that
the
transition rates between two configurations $c$ and $c'$ satisfy
{\it detailed balance\/},
\begin{equation}
\label{eq:detail2}
{ {\cal R}({c'} \rightarrow {c}) } / { {\cal R}({c} \rightarrow {c'}) }
=
\exp{ \left[ - \left({ {\cal H}({c}) -
{\cal H}({c'})} \right) / {k_{\rm B} T} \right] }
\;.
\end{equation}
This result applies to both continuum and discrete systems, and
$\cal H$ may be a classical potential of predetermined form, or the
interaction energies can be calculated ``on the fly" by DFT
\cite{WANG04B}. The sampling can 
be accomplished with a number of different choices of the transition
rates ${\cal R}({c'} \rightarrow {c})$
\cite{LAND00,BROW99A,RIKV02,RIKV02B,RIKV03,PARK04,BUEN04,BUEN06A,BUEN06B,BUEN07},
including Metropolis, Glauber, and heat-bath algorithms.
It is important to note that
the stochastic sequence of configurations generated by an equilibrium
MC algorithm does {\em not\/} generally correspond
to the actual dynamics of the system.

\subsection{Kinetic Monte Carlo}
\label{sec:KMC}

To construct a MC algorithm producing a stochastic path through
configuration space that is a good approximation to the actual time
evolution of the system (in a coarse-grained sense), one can
introduce {\it transition states\/} between the lattice-gas states.
Only then can ``Monte Carlo time,'' measured in MC steps per site
(MCSS) in a lattice-gas simulation, be considered
proportional to ``physical time,'' measured in seconds \cite{HAMA04}.
In a Butler-Volmer approximation \cite{SCHM96,BROW99A},
the free energy of the transition state between lattice-gas
configurations  $c$ and $c'$ is  given by
\begin{equation}
{\cal H}^*\left(c,c'\right) = \Delta +
\left(1-\alpha\right) {\cal H}_{\rm LG}\left(c\right)+
\alpha {\cal H}_{\rm LG}\left(c'\right)
\;,
\end{equation}
where the symmetry constant $\alpha = 1/2$ for diffusion but
may be different for adsorption/desorption \cite{BROW99A}.
The ``bare'' barrier $\Delta$ must be determined by other methods.
These may be {\it ab initio\/} calculations
\cite{WANG04,IGNA98,IGNA97,WANG02,BOGI00},
MD simulations of the diffusion process
on a short time scale as in Sec.~\ref{sec:LI} 
\cite{Allen:92,Haile:92,Rapaport:04,VOTE97,VOTE97B}, 
or comparison of dynamic simulations with experiments \cite{HAMA04}.
The most common choice of transition rate
for KMC in chemical applications is the one-step algorithm
\cite{KANG89,FICH91},
\begin{equation}
{\cal R}\left({c} \rightarrow{c'}\right)
=
\nu_0 \exp \left[ - \left( {\cal H}^*(c,c')
- {\cal H}_{\rm LG}(c)\right)/k_{\rm B} T\right]
\;,
\label{eq:OSD}
\end{equation}
where $\nu_0$ is an attempt frequency (often of the order of a
phonon frequency ($10^9$ -- $10^{13}$~Hz), but see Ref.~\cite{HAMA04}
for exceptions) that must be determined by other means. 
As we have shown previously 
\cite{RIKV02,RIKV02B,RIKV03,PARK04,BUEN04,BUEN06A,BUEN06B,BUEN07},
in order to obtain reliable structural information from a KMC
simulation, the transition rates must approximate the real physical
dynamics, which includes using transition states with proper energies.
While the need for correct transition rates may seem obvious, it is
regrettably often ignored in the literature. 
The most difficult barrier to estimate is that for
adsorption/desorption, which requires reorganization of the adparticle's
hydration shell. 

Since the transition rates used in KMC of activated processes
are typically small, simulations
that extend to macroscopic times must use a {\it rejection-free\/}
algorithm,
such as the $n$-fold way \cite{BORT75,GILM76} or one of its
generalizations
\cite{FRAN06,KANG89,GELT98,LUKK98,KOPE98B,NIET99B,NOVO01,ALAN02}.
These algorithms simulate the {\it same\/} Markov process as the
``na{\"\i}ve''
MC approach of proposing and then accepting or rejecting individual
moves. Although they require more bookkeeping (see the Appendix of
Ref.~\cite{FRAN06} for an example),
they avoid the large waste of computer time resulting from rejected moves.

\section{Electrochemical First-order Reversal Curve Simulations}
\label{sec:FORC}

The First-order Reversal Curve (FORC)
method was originally developed to enhance the amount of dynamic
information extracted from magnetic hysteresis experiments
\cite{MAYE86,PIKE99,PIKE03,ROBB05}.
We recently proposed that the method can be further developed as
an extension of traditional CV to study the dynamics of phase
transitions in electrochemical adsorption \cite{HAMA07,HAMA08}.

\begin{figure}[t]
\vspace{-0.01truecm}
\begin{center}
\includegraphics[angle=0,width=.46\textwidth]{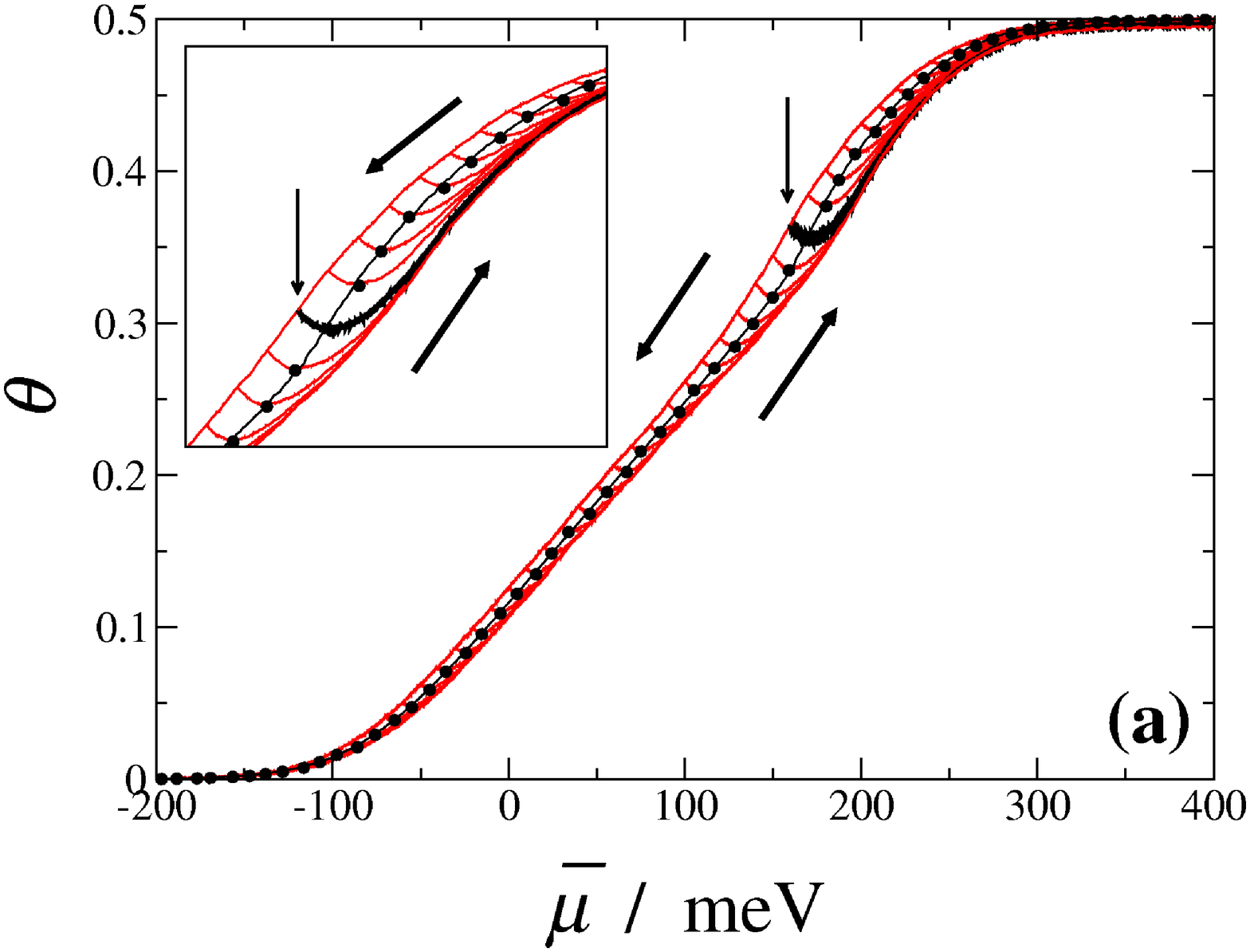}
\hspace{0.01truecm}
\includegraphics[angle=0,width=.52\textwidth]{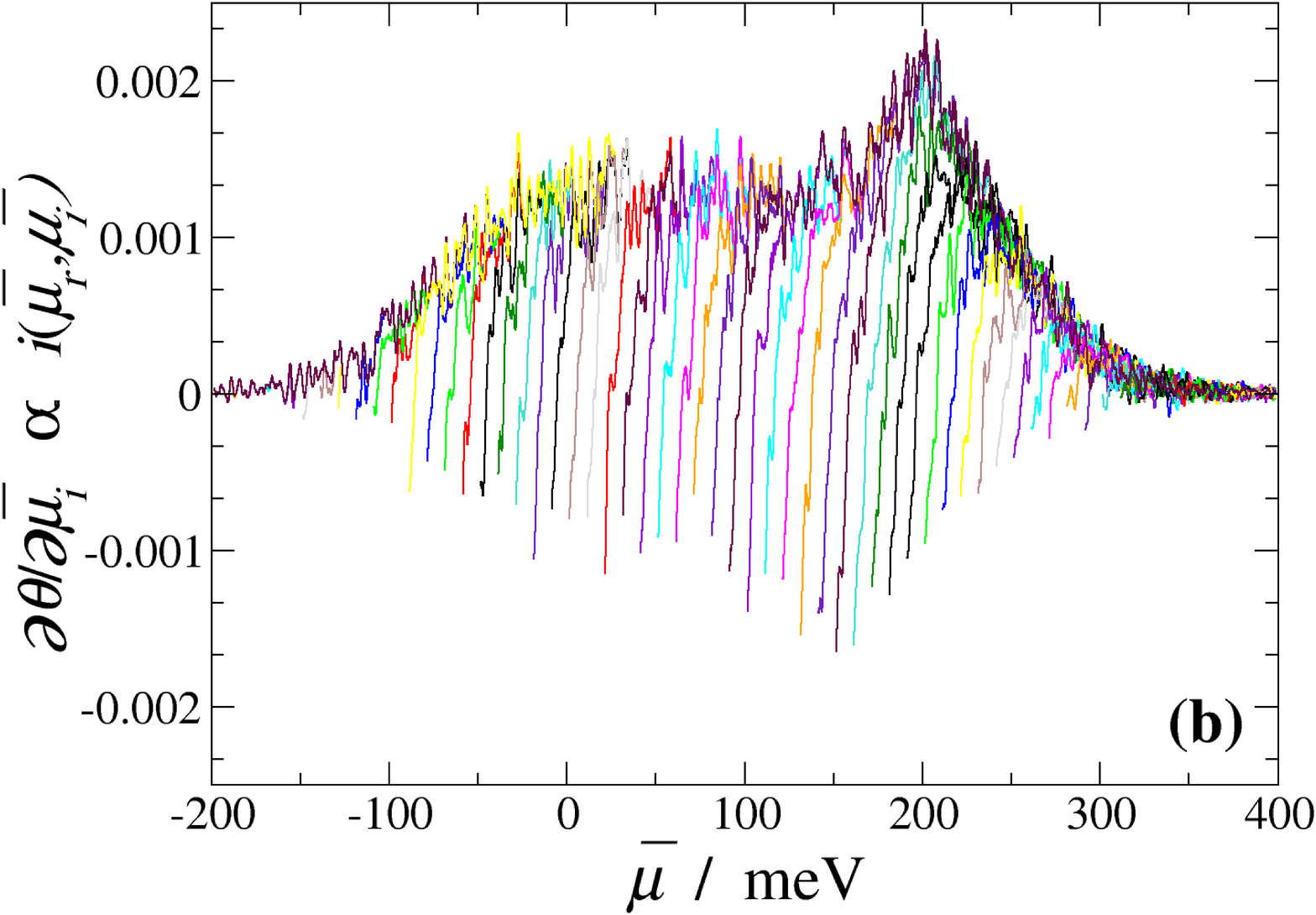}\\
\vspace{0.1truecm}
\includegraphics[angle=0,width=.59\textwidth]{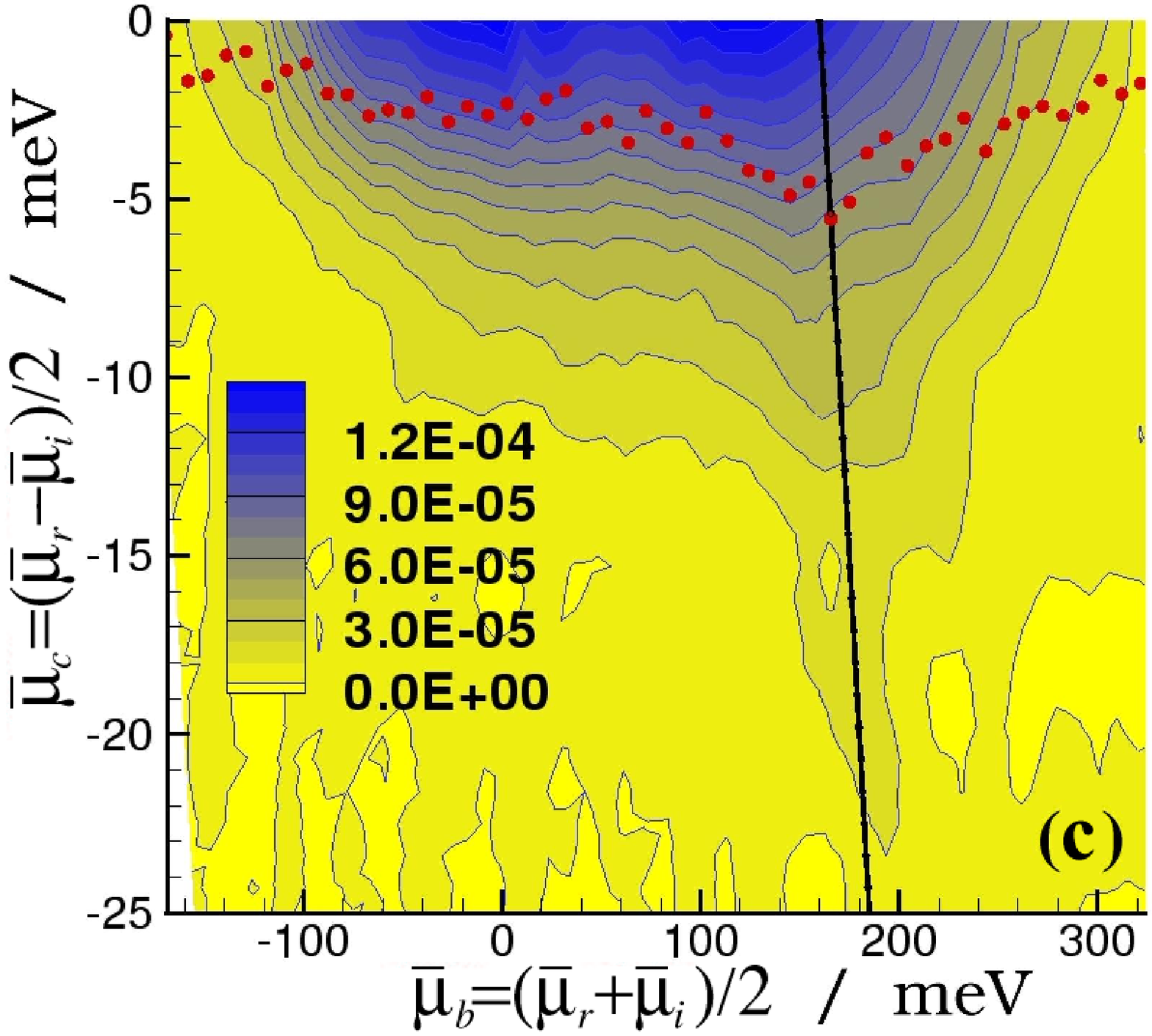}
\end{center}
\caption[]{
({\bf a\/})~Family of FORCs for our model of Br/Ag(100),
corresponding to potential sweeps back and forth across the
continuous phase transition between the disordered and $c(2 \times 2)$ phases.
The bold arrows show the directions of the potential sweeps,
and the vertical arrow indicates $\bar{\mu}_r$ for one of the FORCs.
The bold curve is the FORC whose minimum lies closest to the critical
coverage (shown in more detail in the inset). The thin curve in the
middle is the equilibrium isotherm.
({\bf b\/})~Voltammetric currents corresponding to the FORCs in (a).
({\bf c\/})~Contour plot of
the FORC distribution $\rho$, corresponding to the FORCs
in (a). The jagged curve of dots in the upper part of the diagram
corresponds to the minima of the positive-going curves in (a). The
area above the curve corresponds to desorption, and the area below
it to adsorption. The slanted, straight line corresponds to the
bold curve in (a). After Ref.~\protect\cite{HAMA07}.
}
\label{fig:FORC}
%\vspace*{-0.40cm}
\end{figure}

This electrochemical FORC (EC-FORC) method consists of
saturating the adsorbate coverage $\Theta$ in a strong positive
electrochemical potential $\bar{\mu}$ and, in each case starting from
saturation, decreasing $\bar{\mu}$ at a constant rate
to a series of progressively
more negative ``reversal potentials'' $\bar{\mu}_r$
(see Fig.~\ref{fig:FORC}(a)).
Subsequently, $\bar{\mu}$ is increased back to the saturating
$\bar{\mu}$ at the same rate. 
(Saturation at negative potentials with reversal potentials in the
positive range is also possible.) The method is
thus a simple generalization of the
standard CV method, in which the negative return
potential is decreased for each cycle. This produces a family of
FORCs, $\Theta (\bar{\mu}_r, \bar{\mu}_i)$, where
$\bar{\mu}_i$ is the instantaneous
potential during the increase back toward saturation.
In CV experiments, one actually records the corresponding family of 
voltammetric currents,
\begin{equation}
i(\bar{\mu}_r,\bar{\mu}_i)
=
- \gamma e \frac{{\rm d} \bar{\mu}_i}{{\rm d} t}
\frac{\partial \Theta(\bar{\mu}_r,\bar{\mu}_i)}{\partial \bar{\mu}_i}
\;,
\label{eq:curr}
\end{equation}
where $\gamma$ is the electrosorption valency and $e$ is the elementary
charge (see Fig.~\ref{fig:FORC}(b)).

The next step in extracting dynamical information from the FORCs or
the corresponding currents is to calculate the {\it FORC distribution\/},
\begin{equation}
\rho=- \frac{1}{2} \frac{\partial^2 \Theta}{\partial\bar{\mu}_r\,
\partial\bar{\mu}_i}
=
\frac{1}{2 \gamma e ( {\rm d} \bar{\mu}_i / {\rm d} t )}
\frac{\partial i(\bar{\mu}_r,\bar{\mu}_i)}{\partial \bar{\mu}_r}
\;.
\label{eq:rho}
\end{equation}
This is shown in Fig.~\ref{fig:FORC}(c) 
as a contour plot commonly known as a {\it FORC diagram\/}  
in terms of the more convenient
variables $\bar{\mu}_b = (\bar{\mu}_r + \bar{\mu}_i)/2$
and $\bar{\mu}_c = (\bar{\mu}_r - \bar{\mu}_i)/2$ \cite{HAMA07,PIKE99}.
Geometrically, $\rho$ is proportional to the vertical distance between
adjacent current traces.

To our knowledge, the data for our model of Br/Ag(100)
\cite{HAMA03,HAMA04,RIKV07,MITC00A,MITC00C},
which are shown in Fig.~\ref{fig:FORC}, are the first FORC
predictions for a continuous phase transition. All three
panels are significantly different from the corresponding data for a
discontinuous transition, such as seen in underpotential deposition
(UPD). In particular, the FORC distribution for a discontinuous
transition contains a negative region, while this does not appear for
continuous transitions. (See details in \cite{HAMA07,HAMA08}.)
Closely related to this negative region is an extremum of the current
density during the return scan \cite{FLET83}.
EC-FORC analysis should 
be a useful and valuable method to distinguish between 
continuous and discontinuous phase transitions in experiments.

\section{Conclusion}
\label{sec:CONC}

In this chapter we have presented some applications of
the statistical-mechanics based computer-simulation methods of
Molecular Dynamics and equilibrium and kinetic Monte Carlo simulations
complemented by quantum-mechanical density functional theory
calculations of interaction energies. These include both
highly technologically-oriented applications to Lithium-battery technology,
and basic-science investigations into adsorption on single-crystal 
electrodes. Our hope is that these
examples and the list of references will encourage other workers in
surface electrochemistry to take advantage of the recent spectacular advances
in computational power and algorithmic sophistication to study  
ever-more detailed and accurate models of processes at solid-liquid interfaces.

\section*{Acknowledgements}

This work was supported by U.S.\  National Science Foundation Grant
No.\  DMR-0802288 (Florida State University) 
%and DMR-044405 (Florida State University and Mississippi State University) 
and DMR-0509104 (Clarkson University) and by ABSL Power Solutions, Inc.\  
award No.\  W15P7T06CP408.

%
%
% BibTeX users please use
%\bibliographystyle{prsty}
%\bibliography{elchem.bib,proposal_PRF.bib,md.bib,metastab.bib}
%\bibliography{elchem.bib}
%

%\index{ }
% Use the \index{} command to code your index words

%\index{}
\end{document}